\documentclass[aps,prb,preprint,superscriptaddress,longbibliography]{revtex4-1}

\usepackage{amsmath,amssymb}
\usepackage{graphicx}
\usepackage{epstopdf}
\usepackage{bm}
\usepackage{color}

\usepackage[pagewise]{lineno}

\usepackage{multibib}

\def\be{\begin{equation}}
\def\ee{\end{equation}}
\def\ba{\begin{eqnarray}}
\def\ea{\end{eqnarray}}

\newcommand{\fecose}  {Fe$_{1-x}$Co$_x$Se}
\newcommand{\feses}  {FeSe$_{1-y}$S$_y$}

\newcommand{\se}	{$^{77}$Se}

\newcommand{\slr} 	{$T_1^{-1}$}
\newcommand{\slrt} 	{$(T_1T)^{-1}$}

\newcommand{\tnem} {$T_\text{nem}$}
\newcommand{\tc} {$T_\text{c}$}

\newcommand{\bc}[1]{\textbf{\sffamily #1}}

\begin{document}

\title{Separate tuning of nematicity and spin fluctuations to unravel the origin of superconductivity in FeSe}
\author{Seung-Ho Baek}
\email[Corresponding author: ]{sbaek.fu@gmail.com}
\affiliation{Department of Physics, Changwon National University, Changwon 51139, Korea}
\author{Jong Mok Ok}
\affiliation{Department of Physics, Pohang University of Science and
Technology, Pohang 790-784, Korea}
\affiliation{Center for Artificial Low Dimensional Electronic Systems, Institute of Basic Science, Pohang 790-784, Korea} 
\author{Jun Sung Kim}
\affiliation{Department of Physics, Pohang University of Science and
	Technology, Pohang 790-784, Korea}
\affiliation{Center for Artificial Low Dimensional Electronic Systems, Institute of Basic Science, Pohang 790-784, Korea}
\author{Saicharan Aswartham}
\affiliation{IFW Dresden, Helmholtzstr. 20, 01069 Dresden, Germany}
\author{Igor Morozov}
\affiliation{IFW Dresden, Helmholtzstr. 20, 01069 Dresden, Germany}
\affiliation{Moscow State University, Moscow 119991, Russia}
\author{Dmitriy Chareev}
\affiliation{Institute of Experimental Mineralogy, Russian Academy of Sciences, 142432, Russia}
\affiliation{Ural Federal University, Ekaterinburg 620002, Russia}
\affiliation{Kazan Federal University, Kazan 420008, Russia}
\author{Takahiro Urata}
\affiliation{Department of Physics, Graduate School of Science, Tohoku University, Sendai 980-8578, Japan}
\affiliation{Department of Materials Physics, Nagoya University, Chikusa-ku, Nagoya 464-8603, Japan}
\author{Katsumi Tanigaki}
\affiliation{Department of Physics, Graduate School of Science, Tohoku University, Sendai 980-8578, Japan}
\author{Yoichi Tanabe}
\affiliation{Department of Physics, Graduate School of Science, Tohoku University, Sendai 980-8578, Japan}
\affiliation{Department of Science, Okayama University of Science, Okayama 700-0005, Japan}
\author{Bernd B\"uchner}
\affiliation{IFW Dresden, Helmholtzstr. 20, 01069 Dresden, Germany}
\affiliation{Department of Physics, Technische Universit\"at Dresden, 01062 Dresden, Germany}
\author{Dmitri V. Efremov} 
\affiliation{IFW Dresden, Helmholtzstr. 20, 01069 Dresden, Germany}
\date{\today}

\begin{abstract}
\bf 
The interplay of orbital and spin degrees of freedom is the fundamental characteristic in numerous condensed matter phenomena,  including high temperature superconductivity, quantum spin liquids, and topological semimetals. In iron-based superconductors (FeSCs), this causes  superconductivity to emerge in the vicinity of two other instabilities: nematic and magnetic. Unveiling the mutual relationship among nematic order, spin fluctuations, and superconductivity has been a major challenge for research in FeSCs, but it is still controversial. 
Here, by carrying out \se\ nuclear magnetic
	resonance (NMR) measurements on FeSe single crystals, doped by cobalt and sulfur that serve as control parameters, we demonstrate that the superconducting transition temperature \tc\ increases in proportion to the strength of spin fluctuations, while it is independent of the nematic transition temperature \tnem. 
Our observation therefore directly implies that
superconductivity in FeSe is essentially driven by spin fluctuations in the intermediate coupling regime, while nematic fluctuations have a marginal impact on \tc. 

\end{abstract}

\maketitle

\subsection*{Introduction}
In correlated Fermi fluids, nematicity refers to the state in which rotational symmetry is spontaneously broken, while time-reversal invariance is preserved, and consequently, the symmetry of the crystal changes from tetragonal to orthorhombic.\cite{fradkin10a}
An important aspect in iron-based superconductors (FeSCs) is the propensity for nematic ordering, which is usually followed by a spin-density-wave (SDW) transition, near a superconducting (SC) dome.\cite{fernandes14,bohmer15a,si16}  Regardless of the origin of nematicity that is still under debate,\cite{yamakawa16,chubukov16} this raises the fundamental issue of whether superconductivity in FeSCs is closely related to nematicity\cite{kuo16,wang18} or magnetism\cite{wang11d,dai15} or both.\cite{she18} To address this issue, it is much desirable to separate nematic order from magnetic one. In this respect, FeSe has been a key platform for studying the origin of nematicity and its role in superconductivity,\cite{bohmer18} as it exhibits nematic and SC orders at well separated temperatures, $T_\text{nem}\sim 90$ K and $T_\text{c}\sim 9$ K, respectively, without involving magnetic order. 
Numerous recent studies in FeSe show that nematicity causes the strongly anisotropic SC gap symmetry,\cite{xu16,hashimoto18,liu18,kushnirenko18} and further discuss that nematic fluctuations might play an important role for the superconducting pairing mechanism.\cite{hosoi16,matsuura17}  On the other hand,  the leading role of spin fluctuations (SFs) in the SC mechanism of FeSe, as in other FeSCs whose parent materials magnetically orders, has been also proposed in the literature.\cite{imai09, sun17,wiecki18, rhodes18} In this spin fluctuation-mediated pairing scenario, the subsequent question arises whether weak or strong coupling approach is appropriate to establish theory of superconductivity in FeSCs. It is quite interesting to note that recent NMR studies of FeSe under high pressure reveal the persistence of local nematicity at temperatures far above \tnem, which suggests a correlation between local nematicity and magnetism.\cite{wang17,wiecki17} Another interesting observation by NMR is the unusual suppression of \slrt\ at optimal pressure,\cite{kuwayama19} suggesting that the interplay of SFs and superconductivity may undergo a critical change with high pressure.
 
 As a system undergoes a nematic transition ($C_4\rightarrow C_2$), two nematic domains are naturally formed below \tnem, still preserving the $C_4$ symmetry on average. Accordingly, it is usually required to detwin nematic domains, for example, by an external strain to study nematicity. As a local probe in real space, on the other hand, NMR is uniquely capable of observing the two nematic domains at the same time. Indeed,
it has been established that the splitting of the NMR line in FeSCs at an external field $H$ applied along the crystallographic $a$ axis represents the nematic order parameter and its onset temperature corresponds to the nematic transition temperature \tnem\  (Ref. \onlinecite{baek15,baek16,ok18}, see Fig. 1c). In order to investigate whether and how nematicity is related to superconductivity, we measured the \se\ line splitting for $H_{\parallel a}$ in \feses\ and \fecose\ single crystals.  In general, it is considered that substituting isovalent S for Se is equivalent to the application of (negative) chemical pressure, and Co substituted for Fe supplies an additional electron and also plays as a paramagnetic impurity. Therefore, a systematic NMR study on the two different doped systems may enable a full understanding of the relationship between nematicity, magnetism, and superconductivity. 

\subsection*{Results and Discussion}
Figure 1 shows how the temperature dependence of the \se\ NMR spectrum in FeSe is modified as a function of $x$ in \fecose\ (Figs. 1a and 1b) and as a function of $y$ in \feses\ (Figs. 1d and 1e). For \feses, we find that  the onset temperature of the line splitting or \tnem\ is gradually suppressed,  consistent with previous studies.\cite{hosoi16,wiecki18} For \fecose, however, \tnem\ hardly changes for  $x=0.018$. Upon further doping to slightly higher $x=0.025$, the \se\ line becomes significantly broad, making difficult to identify the onset of the line splitting. 
 The much larger \se\ line broadening for Co doping than for S doping is well understood because Co has a strong influence on Fe moments as a nonmagnetic impurity. We notice, however, that the \se\ line broadening is not simply proportional to the concentration of Co dopants, but rather it appears to increase drastically above $x\sim 0.025$. In fact, for $x=0.036$, it is not possible to observe the line splitting anymore, because the linewidth is much larger than the nematic splitting (see supplementary Fig. 1). On the other hand, the \se\ linewidth is proportional to both S and Co dopants similarly, as long as Co-doping is smaller than 2.5\%, as shown in supplementary Fig. 2.
 This suggests that doped Co impurities beyond $\sim 2.5$\% of Fe sites causes a strong disorder effect on the correlation between Fe spins, indicating the existence of a critical doping level above which the magnetic correlation length becomes sufficiently long to induce a short-range exchange. 

Although \tnem\ cannot be accurately determined for $x=0.025$ due to the large line broadening in the nematic state, we clearly observed the line splitting below 80 K, as shown in Fig. 1a. While this puts a lower limit of \tnem, the fitting analysis of \se\ spectra in  Fig. 1a also suggests that the line splitting seems to persist even up to 100 K (see vertical bars). (The detailed Knight shift data as a function of temperature and doping are shown in supplementary Fig. 3.) This suggests that Co impurities may induce a spatial distribution of \tnem\ in the temperature range, $80\leq T_\text{nem}\leq 100$ K.  Regardless of details, \tnem\ is marginally suppressed by Co doping within the doping range investigated, as shown in Fig. 2a. Note that the Co doping range investigated is very narrow, and thus we are unable to argue whether \tnem\ remains a constant at higher Co-doping. In any case, Figs. 2a and 2b reveal that \tnem\ and \tc\ are clearly decoupled. 

Interestingly, the split \se\ lines below 80 K for $x=0.025$ is notably anisotropic, i.e., the peak for the lower frequency side is broader than that for the higher frequency. The origin of the anisotropic line shape is unclear, but we note that the  similar anisotropic \se\ line shape is also observed  at $T<20$ K for $x=0.018$ (see Fig. 1b). This implies that magnetic inhomogeneity, which otherwise appears at low temperatures, prevails at higher temperatures with higher Co doping. 

Contrasting sharply with the weak dependence of nematicity on both  S and Co dopants, our susceptibility measurements reveal that superconductivity is strongly dependent only on Co dopants. That is, \tc\ is rapidly suppressed by small Co doping, whereas it is robust with regard to S doping, as shown in Figs. 2a and 2b, being consistent with previous studies.\cite{urata16,abdel-hafiez16}
The very different behavior of \tnem\ and \tc\ with doping indicates that nematic and superconducting orders are not directly coupled,\cite{coldea16, massat18} raising a strong question as to whether nematicity and superconductivity are closely related.\cite{lederer15,hosoi16,kuo16,wang18}

Having established the lack of a coupling of the nematic and superconducting transition temperatures, we now discuss the role of SFs for superconductivity. 
For probing low energy SFs, we measured the spin-lattice relaxation rate, \slr, as the quantity \slrt\ is a measure of SFs at very low energy:
\begin{equation}
(T_1 T)^{-1}  = \gamma_n^2  \lim_{\omega\to 0} 
\sum_{\mathbf{k}} A^2(\mathbf{k}) \frac{\chi''(\mathbf{k}, \omega)}{\omega},
\end{equation}
where $\chi''(\mathbf{k},\omega)$ is  the imaginary part of the dynamic susceptibility at momentum $\mathbf{k}$ and frequency $\omega$,  $\gamma_n$ is the
nuclear gyromagnetic ratio, and $A(\mathbf{k})$ is the structure
factor of the hyperfine interaction.
Figures 2c and 2d show \slrt\ as a function of Co and S doping, respectively, at $H_{\parallel a}=9$ T. The data for the undoped FeSe crystal was taken from ref.~\onlinecite{baek15}. 
With increasing Co doping $x$ in \fecose, \slrt\ or SFs above \tc\ is rapidly suppressed, which is in exact parallel with the suppression of \tc, as shown in Fig. 2a. Note that for $x=0.035$ superconductivity is completely absent, and correspondingly SFs are not enhanced at all at low temperatures. 
On the other hand,   \slrt\ above \tc\  is unchanged with increasing S doping $y$ in \feses\ up to $y=0.1$, as precisely \tc\ does (see Fig. 2b).   From the data presented in Figure 2, therefore, one sees that \tc\ depends only on the strength of spin fluctuations,  but not on \tnem. (At larger S-doping near $y=0.2$, it was reported that both \slrt\ and \tc\ are strongly suppressed in such a way that the correlation between SFs and \tc\  persists,\cite{wiecki18} somewhat similar to the behavior in Co-doped samples.)

For further quantitative information on how SFs is related to \tc, we adopt a spin fluctuation model in the Eliashberg formalism,\cite{allen82} or Millis-Monien-Pines (MMP) model.\cite{millis90} For this, we separate out the enhancement of \slrt\ that is solely associated with SFs from the data shown in Fig. 2c. Noting that \slrt\ for the non-superconducting sample ($x=0.035$) approaches a constant, $(T_1T)^{-1}_0\equiv \Gamma_0$, without any enhancement at zero temperature,  one may define the strength of SFs $\Gamma$ for $H_{\parallel a}$  from the $(T_1T)^{-1} (x)$ values just above \tc:  
\begin{equation}
	\Gamma(x)\equiv (T_1T)^{-1} (x)|_{T_\text{c}}-\Gamma_0.
\end{equation}

While the MMP model indicates that $\Gamma$ is proportional to the square of the correlation length,\cite{millis90} $\xi^2(T)$,    
the estimation of the low energy part of the Eliashberg bosonic spectral function suggests that the coupling constant $\lambda$ is proportional to $\xi$, i.e., $\sqrt{\Gamma}$. 
As it was analyzed by Radtke \textit{et al.}\cite{radtke92} and Popovich \textit{et al.}\cite{popovich10} the direct use of the MMP-spectrum gives overestimation of \tc\ and the gap function due to a long tail at high energies of the bosonic spectral functions $\sim 1/\omega$. To cure the problem it was proposed to introduce a cut-off or calculate the bosonic self-energy at high energies.\cite{abanov03} For simplicity we use the approach of cut-off proposed in ref.~\onlinecite{popovich10}. A detailed procedure of the calculation is described in Supplementary Note 1.

The plot of \tc\ vs.~$\sqrt{\Gamma}$  is shown in Fig. 3. 
The solid curve is a theoretical calculation (\tc\ vs.~$\lambda \propto \sqrt{\Gamma}$) based on the Eliashberg theory in which electron correlation effects are substantial.  The good agreement of our theory with the experimental data evidences that the magnetic scenario for superconductivity in which Cooper pairing is mediated by spin fluctuations applies to FeSe, and it is likely a universal superconducting mechanism among FeSCs. 

Based on our NMR finding that \tc\ relies only on SFs,  the seeming relevance of nematicity with superconductivity may be simply due to the closeness with magnetism, rather than to superconductivity itself. It should be noted that the strongly anisotropic gap structure\cite{xu16,hashimoto18,liu18,rhodes18,kushnirenko18} observed in FeSe may be a natural consequence of the presence of nematicity within the superconducting state. It is because nematicity involves the splitting of $d_{xz}$ and $d_{yz}$ orbitals which should have an inevitable influence on the gap symmetry. However, \tc\ itself is not necessarily affected by nematicity.\cite{kang18} 
Nevertheless, nematicity may be considered as an important barometer for superconductivity in FeSCs, as it is strongly coupled to magnetism\cite{matsuura17} which in turn directly correlates with superconductivity.

\subsection*{Methods}

\paragraph*{Crystal growth and characterization.} The growth of \fecose\ and \feses\ single crystals was performed by 
using the KCl - AlCl$_3$ flux technique in permanent T-gradient in accordance 
with refs.~\onlinecite{chareev13,chareev18}. All preliminary operations for 
the preparation of the reaction mixture were carried out in a dry box with a 
residual pressure of O$_2$ and H$_2$O not higher than 0.1 ppm. At the first 
stage, polycrystalline samples of the composition \fecose\ and \feses\ were 
obtained. For this, Fe, Co, S and Se powders were carefully ground in a mortar 
in the appropriate ratio, and then annealed in evacuated quartz ampoules at 
420 $^\circ$C for few days. In the second stage, 0.5 g of the prepared sample 
was placed on the bottom of a thick-walled ampoule, and then the mixture of 
AlCl$_3$ and KCl in a molar ratio of AlCl$_3$:KCl = 2:1 is added to the 
ampule, after that the ampule was evacuated and sealed. The sealed ampoule 
with polycrystalline sample of Fe-Co-Se-S loads was placed in a horizontal 
2-zone furnace and heated for five weeks in such a way that, the hot zone 
temperature was set to 420 $^\circ$C and the cold zone temperature was set to 
370 $^\circ$C. After five weeks, the furnace was turned off and the ampoule 
was removed from the furnace. Next, the ampoule was cut and the single 
crystals from cold zone were separated from the flux by dissolving it in 
water. The single crystals obtained were thin square plates with metallic 
luster. The single crystals were grown with platelet like morphology and were 
characterized by SEM/EDX for compositional analysis. 

\paragraph*{Nuclear magnetic resonance.} \se\ (nuclear spin $I=1/2$) NMR was carried
out in undoped and doped FeSe single
crystals at an external magnetic field and in
the range of temperature 4.2 -- 160 K.
The samples were oriented using a
goniometer for the accurate alignment along the external field. The \se\ NMR
spectra were acquired by a standard spin-echo technique with a typical $\pi/2$
pulse length 2--3 $\mu$s.
The nuclear spin-lattice relaxation rate \slr\ was obtained by fitting the
recovery of the nuclear magnetization $M(t)$ after a saturating pulse to
following fitting function, 
\begin{equation}
\nonumber
	 1-M(t)/M(\infty) = A\exp(-t/T_1)
\end{equation}
where $A$ is a fitting parameter that is ideally unity. 

\paragraph*{Determination of \tc\ and \tnem.}
The superconducting transition temperature \tc\ was determined from magnetic susceptibility ($\chi$) measurements by comparing field-cooled and zero-field cooled data, while we obtained the nematic transition temperature \tnem\ by measuring the temperature at which the \se\ NMR line splits (see Fig. 1). Due to the weakness of the signal intensity, we were unable to determine \tc\ by \slrt\ measurements except the undoped FeSe sample. This could give an error in extracting spin fluctuations just above \tc, $\Gamma$, which was reflected in an experimental error indicated in Fig. 3.

\let\origdescription\description
\renewenvironment{description}{
  \setlength{\leftmargini}{0em}
  \origdescription
  \setlength{\itemindent}{0em}
  \setlength{\labelsep}{\textwidth}
}
{\endlist}

\begin{description}
\item[Data Availability] The data that support the findings of this study are available from the corresponding author (S.-H. Baek).
\item[Acknowledgments] 
We acknowledge A. Chubukov for useful discussions. 
S.H.B has been supported by the Deutsche
		Forschungsgemeinschaft (Germany) via DFG Research Grants BA 4927/2-1 and by the National Research Foundation of Korea (NRF-2019R1F1A1057463). D.V.E., B.B., I.M., and S.A. were supported by RSF-DFG project (no. 19-43-04129, BU887/25-1, EF86/7-1). D.V.E and I.M. were also supported  by VW foundation in the frame of the VW Trilateral Initiative.
		S.A. acknowledges financial support from Deutsche Forschungsgemeinschaft (DFG) via Grant No. DFG AS 523/4-1.
		The work at POSTECH was supported by Institute for Basic Science  (no. IBS-R014-D1) and also by the National Research Foundation (NRF) of Korea through the SRC (no. 2018R1A5A6075964) and the Max Planck-POSTECH Center for Complex Phase Materials in Korea (MPK) (no. 2016K1A4A4A01922028).
		The work of DACh was supported by the program 211 of the Russian Federation Government, agreement No. 02.A03.21.0006, by the Russian Government Program of Competitive Growth of Kazan Federal University.
\item[Competing Interests] The authors declare no
	competing financial or non-financial interests.				
\item[Author Contributions] SHB and BB have proposed and initiated the
	project.  JMO, JSK, SA, IM, DC, TU, KT, and YT have grown single crystals and characterized
		transport and superconducting properties. SHB and JMO performed NMR
		measurements and analyzed data;
    SHB, DVE, and BB participated in writing of the manuscript. All authors
		discussed the results and commented on the manuscript.
\item[Additional information] Correspondence and requests for materials should be
	addressed to S.-H. Baek~(email: sbaek.fu@gmail.com).
\end{description}

\bibliographystyle{naturemag}



\pagebreak

\begin{figure*}
\centering
\includegraphics[width=\linewidth]{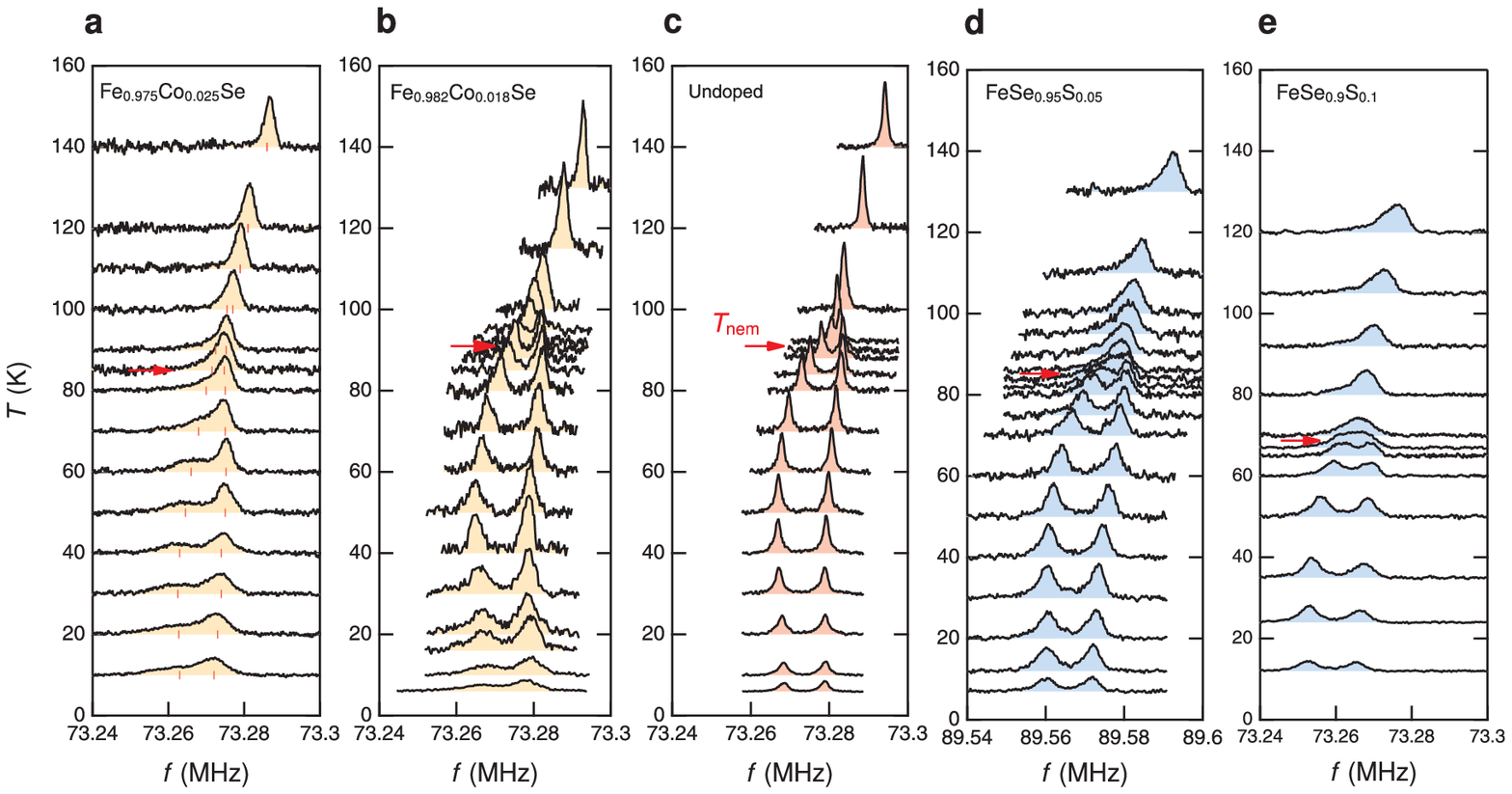}
\caption{\se\ NMR spectra in undoped and doped FeSe single crystals for $H \parallel a$.
	\bc{a}-\bc{b}, Temperature dependence of \se\ spectrum of \fecose. For $x=0.018$ (\bc{b}), the \se\ spectrum shows a very similar behavior as the undoped one, except a moderate line broadening. For a slightly larger doping, $x=0.025$ (\bc{a}), the \se\ line undergoes a considerable line broadening. While the splitting of the two \se\ lines were clearly identified at low temperatures (vertical bars), the onset of the splitting is not well defined, being ascribed to local disorder.
	\bc{c}, Temperature dependence of \se\ spectrum for undoped FeSe. 
	\bc{d}-\bc{e}, Temperature dependence of \se\ spectrum of \feses\ for $y=0.05$ and 0.1, respectively. \tnem\ is progressively suppressed with increasing S doping.
}
\label{spec}
\end{figure*}

\pagebreak

\begin{figure*}
\centering
\includegraphics[width=\linewidth]{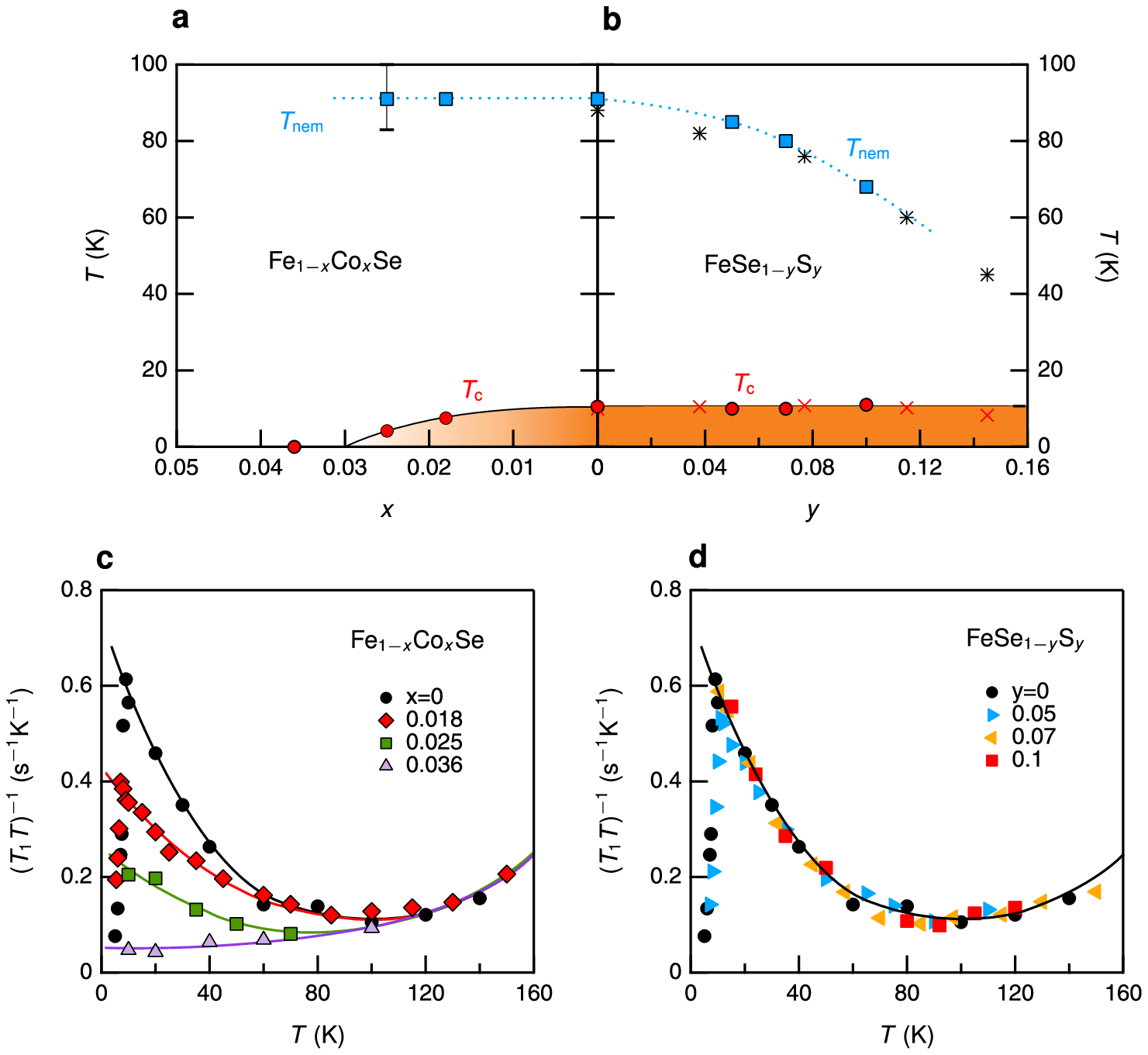}
\caption{Doping dependence of spin fluctuations for $H\parallel a$ in FeSe.
	\bc{a} and \bc{b}, The phase diagrams of \fecose\ and \feses, respectively.  For Co doping, \tnem\ is hardly influenced, but \tc\ is strongly suppressed. In contrast, for S doping, \tnem\ is suppressed with increasing $y$ and \tc\ remains nearly the same. The asterisk(*) and cross ($\times$) symbols are the \tnem\ and \tc\ data, respectively, extracted  from Ref. \onlinecite{hosoi16}. 
	\bc{c}, The spin-lattice relaxation rate divided by
	temperature, \slrt\ which measures spin fluctuations, as a function of temperature and Co-doping $x$ in \fecose. The enhancement of \slrt\ at low temperatures is progressively suppressed with increasing $x$ (see Fig. 3). The solid lines are guides to the eyes. 
	\bc{d}, \slrt\ as a function of temperature and S-doping $y$ in \feses. Spin fluctuations are unchanged with increasing S doping $y$, being consistent with \tc\ that is nearly independent of $y$. 
}
\label{t1t}
\end{figure*}

\begin{figure*}
\centering
\includegraphics[width=0.5\linewidth]{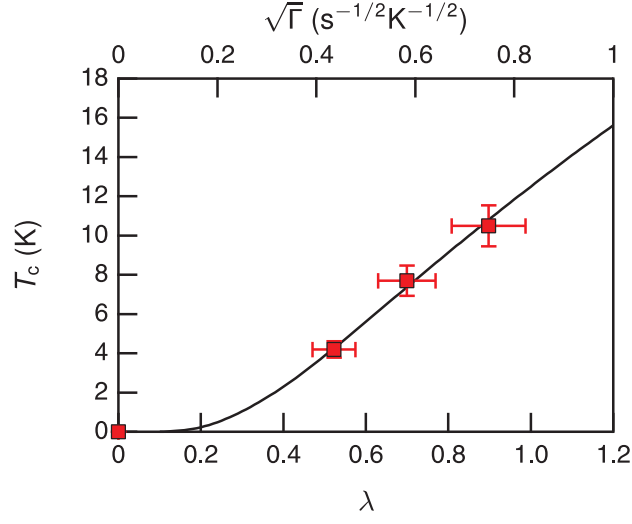}
\caption{Superconducting transition temperature \tc\ vs.~$\sqrt{\Gamma}$, where $\Gamma$ is the strength of spin fluctuations just above \tc\ for $H_{\parallel a}$ in Co-doped FeSe single crystals. The solid line represents our theory of \tc\ vs.~the coupling constant $\lambda \propto \sqrt{\Gamma}$ (see text). 
The error bars represent the uncertainty in determining \tc\ and $\Gamma$. }
\label{SF}
\end{figure*}

\pagebreak
\clearpage

\setcounter{figure}{0}
\renewcommand{\figurename}{{\bf Supplementary Figure}}
\renewcommand{\thefigure}{{\bf \arabic{figure}}}
\renewcommand\thesubsection{Supplementary Note \arabic{subsection}}
\tolerance=400
\emergencystretch=10pt

	{\bf Supplementary Material to ``Separate tuning of nematicity and spin fluctuations to unravel the origin of superconductivity in FeSe"}

\subsection{Superconducting critical temperature in the strong coupling approximation.}	

The superconducting critical temperature  $T_c$ in the framework of the Eliashberg theory is given by the solution of  the system of the linearized
Eliashberg equations:\cite{allen82,efremov17}
		\begin{align}
		Z_{\alpha n} &= 1 + T_c\sum\limits_{\omega_{n}^{\prime},\beta
		}\lambda_{\alpha\beta}^{z}(n-n^{\prime}) \frac{\mathrm{sign}(\omega_{n'}) }{\omega_n},
		\label{eq.tc1} 
		\\
		Z_{\alpha n}\Delta_{\alpha n} &= T_c\sum\limits_{\omega_{n}^{\prime},\beta
		}\lambda_{\alpha\beta}^{\phi}(n-n^{\prime}) \frac{\Delta_{\beta n'}}{\omega_{n'}}\label{eq.tc2} .
		\end{align}
where $\omega_n = \pi T_c(2 n +1)$ are the Matsubara frequencies,  $\Delta_{\alpha n}$ are the gap functions, $Z_{\alpha n}$ are the $Z$ functions, and $\alpha, \beta$ are the band indices.    
The coupling functions 
\be\lambda^{\tilde\phi,z}_{\alpha\beta}(n-n') = 2\lambda^{\tilde\phi,z}_{\alpha\beta} \!\!
\int^{\infty}_{0} d\Omega \frac{\Omega B(\Omega)}
{[(\omega_n-\omega_{n'})^{2}+\Omega^{2}]}
\label{eq.b(omega)}
\ee
are expressed through  the normalized
bosonic spectral function $B(\Omega)$. 
For the sake of simplicity we neglect the intraband  interaction ($\lambda^{\phi,z}_{11} = \lambda^{\phi,z}_{22} =0$).
The interband coupling constants $\lambda^{\tilde\phi}_{\alpha \beta}$ are chosen to be negative (repulsive) due to the prevailing spin-fluctuation mechanism of the electron-electron interaction.\cite{chubukov08,mazin09} The matrix
elements $\lambda_{\alpha\beta}^{z}$ are  positive. For simplicity we use the approximation $\lambda_{\alpha\beta}^{z}=|\lambda_{\alpha\beta}^{\tilde\phi}|\equiv
|\lambda_{\alpha\beta}|$ and  neglects the $\mathbf{k}$-space anisotropy in
the gap functions ${\Delta}_{\alpha n}$.  

\paragraph{Spin-fluctuation model.}
A phenomenological model of spin fluctuation spectra based on NMR data was proposed for nearly antiferromagnetic metals by Millis et al. \cite{millis90} [Millis-Monien-Pines (MMP) model].
They argued  that the low energy dynamical spin susceptibility may be represented in the following form:
\begin{equation}
Im\chi(\mathbf{q},\omega)  = \frac{\chi_0 \Gamma_\text{sf}}{\pi \omega_\text{sf}} \frac{ (\omega/\omega_\text{sf})}{(1 + \xi^2 |\mathbf{q} - \mathbf{Q}|^2 )^2 +(\omega/\omega_\text{sf})^2} \times \Theta(\Omega_\text{c} -|\omega|),
\label{eq.MMP}
\end{equation}
where $Q$ is the nesting vector, $\Omega_c$ is the energy cut-off and 
$$
\omega_\text{sf}= \Gamma_\text{sf} \frac{a^2}{\pi^2 \xi^2},
$$
where $\xi$ is the correlation length, characterizing the the proximity to the antiferromagnetic instability. At the transition point it diverges $\xi^{-1} =0$.   
The static spin susceptibility is denoted as $\chi_0$, $a$ is the lattice constant, and  $\Gamma_\text{sf}$ is the frequency scale characterizing the spin fluctuations.   
Within this model one gets  the spin lattice relaxation rate in the following form at low temperatures (see also ref. \onlinecite{millis92}.):
\begin{equation}
\Gamma=	\frac{1}{T_1 T} = \gamma_g^2 \lim_{\omega\to 0} \sum_{\mathbf{q}} \frac{M^2\chi''(\mathbf{q},\omega)}{\omega} \sim \frac{\chi_0}{\xi^{-2}}.  
\end{equation}  

\paragraph{Boson spectral function.} The main input into the Eliashberg equations Eqs.\eqref{eq.tc1} and \eqref{eq.tc2} is the spectral function of the intermediate bosons $B(\mathbf{q},\omega)$: 
\begin{equation}
\alpha^2 F(\omega) = \frac{\sum_{\mathbf{k,k'}} \delta(\epsilon_k)\delta(\epsilon_{k'})|A(\mathbf{k},\mathbf{k'})|^2B(\mathbf{k-k'},\omega)}{ \sum_{\mathbf{k}}\delta(\epsilon_\mathbf{k})}
\label{eq.alpha2F}
\end{equation}
where $A(\mathbf{k},\mathbf{k'})$ is the matrix element for the scattering an electron in Bloch state $\mathbf{k}$ to $\mathbf{k'}$
and
\begin{equation}
B(\mathbf{q},\Omega) = -\frac{1}{\pi} N(0) Im \chi(\mathbf{q},\Omega)
\end{equation}
is the spectral function of the spin fluctuations normalized by the density  of states at the Fermi level $N(0)$. 
As it was pointed out in refs.~\onlinecite{schuettler96,benfatto09,popovich10}, that use of the MMP spectrum in the Eliashberg equation leads  to overestimation of $T_\text{c}$ and superconducting gaps due to a long tail at high frequencies $\propto 1/\omega$.
Here, we adopt the phenomenological approach proposed by Popovich et al.\cite{popovich10} The low energy part of the bosonic function is given by Eq.~\eqref{eq.MMP} and the function decays fast after a characteristic cut-off energy. The cut-off energy is determined by the band structure and in the leading approximation can be be taken independent on the distance to the quantum critical point. 

Since we consider the leading term, we neglect the momentum dependence of $A = g= \text{const}$. 
Performing the momentum integration in Eq. \eqref{eq.alpha2F}  we get for small $\omega$: 
\begin{equation}
	\alpha^2F(\omega) \approx \frac{g^2 N(0)}{4\pi^2 (\xi/a)} \frac{\omega}{\omega_\text{sf}} \sim g^2 N(0) \frac{ \xi}{ a } \frac{\omega}{\Gamma_\text{sf}}.
\end{equation}

It determines that $\lambda \sim \xi \sim \sqrt{\Gamma}$. In Fig. 3 of the main text we show the experimental and calculated $T_c$ vs. $\lambda$ and $\sqrt{\Gamma}$ correspondingly. The fit gives the scale of the coupling constants. 

\begin{figure}
	\centering
	\includegraphics[width=.5\linewidth]{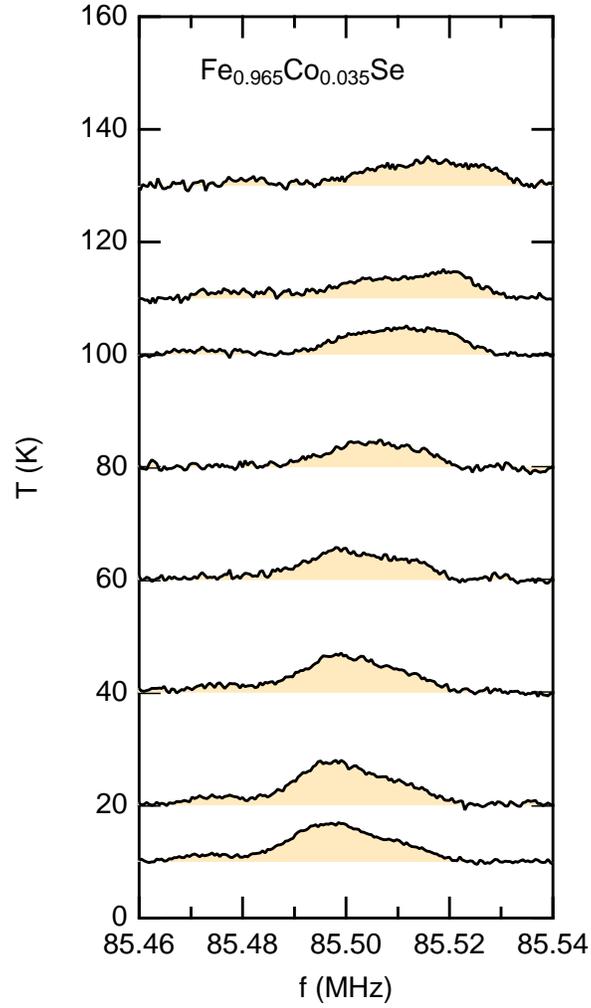}
	\caption{Temperature dependence of \se\ spectrum at 3.6\%Co doping. The full-width-at-half-maximum (FWHM) significantly increased at this doping. Since the splitting of the line due to nematicity is smaller than the FWHM, \tnem\ cannot be identified.}
	\label{S1}
\end{figure}

\begin{figure}[h]
	\centering
	\includegraphics[width=.6\linewidth]{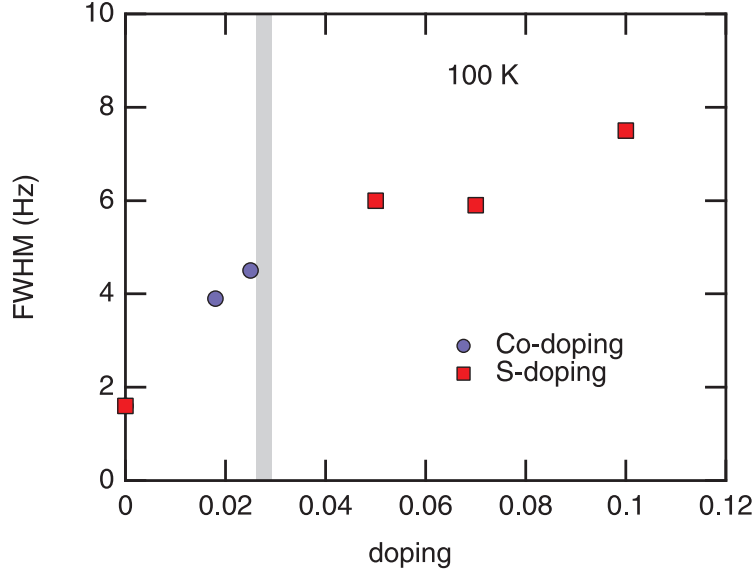}
	\caption{Doping dependence of \se\ full-width-at-half-maximum (FWHM) at 100 K. The gray line denotes the existence of a critical Co doping.}
	\label{S1}
\end{figure}

\begin{figure}[h]
	\centering
	\includegraphics[width=\linewidth]{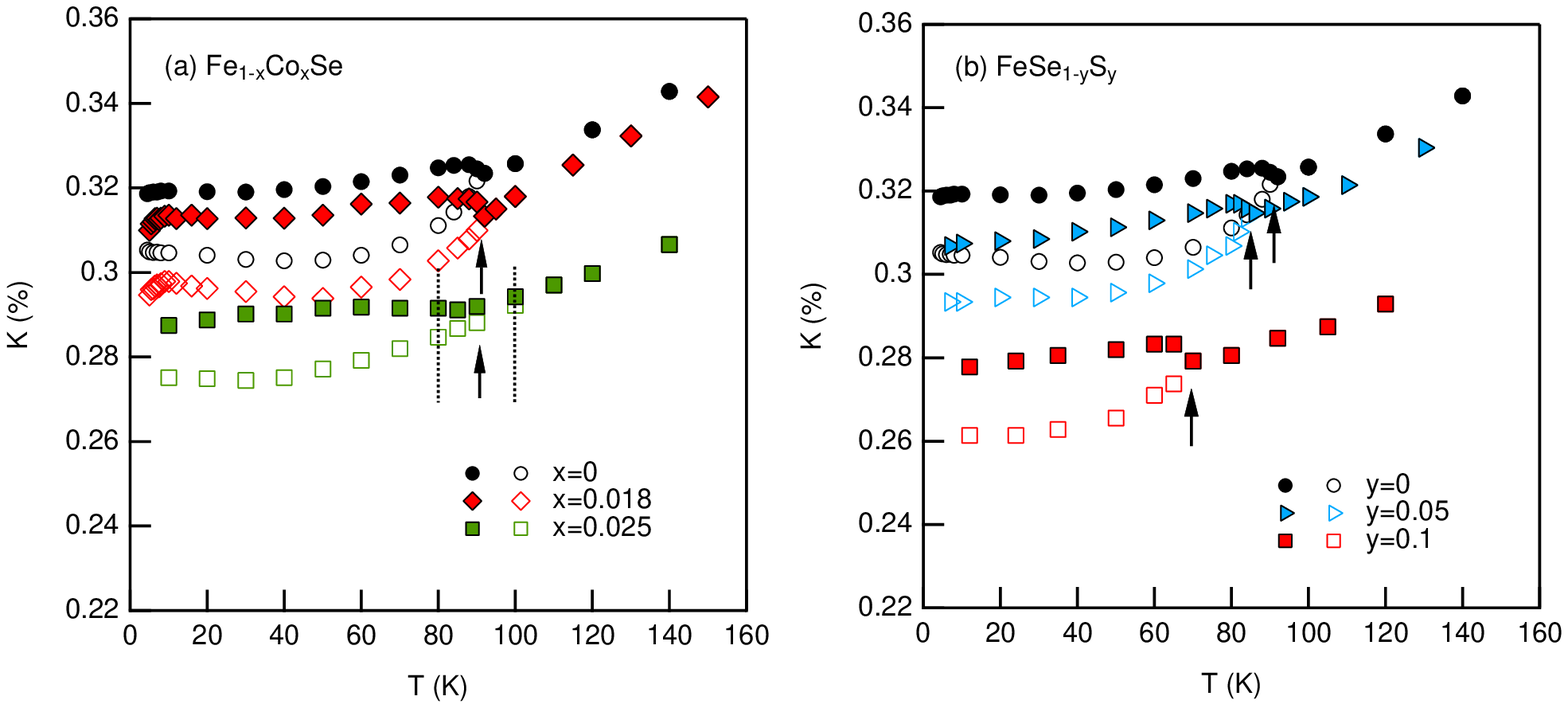}
	\caption{\se\ Knight shift data in (a)\fecose\  (b) \feses. The arrow indicates the nematic transition temperature \tnem. For $x=0.025$ in (a), \tnem\ cannot be identified accurately, but is in the range 80-100 K. It should be noted that the Knight shift splitting $\Delta \mathcal{K}(T)$ is sensitive to an inevitable slight misalignment of the sample with respect to $H$. Therefore, we attribute the small variation of $\Delta \mathcal{K}(T)$ with doping to an experimental error.}
	\label{S1}
\end{figure}

\clearpage

\bibliographystyle{naturemag}

\end{document}